\documentclass[aps,floats,epsf,twocolumn]{revtex4}
 \usepackage{graphics}

\begin{document}

\title{Zero-energy states of massive Dirac equation in magnetic fields}
\author{Igor F. Herbut}

\affiliation{Department of Physics, Simon Fraser University,
 Burnaby, British Columbia, Canada V5A 1S6}

\begin{abstract}
The Dirac equation with a $U(1)$ vortex in the mass-term is solved in the presence of magnetic-like  fields at zero energy. By drawing an analogy to classical mechanics it is shown that the four-component Dirac equation in arbitrary magnetic field always yields one zero-energy state. In the time-reversal preserving, pseudo-magnetic field, however, the number of zero-energy states may depend on the field's profile and sign. Some explicit examples are worked out. Possible implications of these  results for the electrical charge of the vortex and for the behavior of graphene in magnetic field are discussed.
\end{abstract}
\maketitle

\vspace{10pt}

\section{Introduction}

  Relativistic invariance dictates that the spectrum of Dirac's equation extends over both positive and negative values. Even in the condensed-matter context where the invariance is only pseudo-relativistic, the spectrum is often symmetric with respect to the change of the sign of the energy. Particularly interesting then are the states with precisely zero energy, right in the middle of such a spectrum. They have been known to exist in at least two sets of circumstances: when the Dirac electron is subjected to the magnetic field \cite{aharonov}, or when the mass-term forms a kink or a vortex in the configurational space \cite{jackiw}. In either case the number of the zero-energy states is related to some global property of the external potential that acts on the electron, such as the magnetic flux or the vorticity, but is otherwise independent of the details of the configuration. The existence of the macroscopically degenerate manifold of zero-energy states is believed to be responsible for some of the observed quantum Hall effects in graphene in uniform magnetic field \cite{jiang}, for example. Vortex zero-energy states have  been intensely studied theoretically for a number of years, within a scenario for the fractionalization of electric charge \cite{chamon}. In the context of graphene, in particular, the zero-energy states in the core of the vortex of a superconducting order parameter have recently been argued to provide an example of the elusive Majorana fermion \cite{wilczek}, as well as to render the core of the vortex ordered \cite{herbut1}.

    It is less clear if, and what kind of zero-energy states exist in presence of both the vortex in the mass-term in the Dirac equation, and the magnetic fields. This problem arises in several current lines of investigation. In the context of fractionalization, Jackiw and Pi \cite{pi} have shown that the addition of a localized flux of the pseudo, that is time-reversal symmetry preserving, magnetic field does not change the number of zero-energy states, but only modifies their form. The problem of the vortex in the superconducting order in graphene can be mapped onto the same Hamiltonian \cite{herbut1}. The recent observation of the Kosterlitz-Thouless scaling of the resistivity near the  metal-insulator transition in graphene also urgently calls for a better understanding of the vortex structure of Dirac fermions in true, time-reversal symmetry breaking, magnetic field \cite{ong}. At the present time there  are at least two candidates for the possible order parameters in this system with the requisite $U(1)$  symmetry: the N\' eel order parameter, favored by the on-site repulsion between electrons, which acquires an easy plane in  the magnetic field \cite{herbut2}, and the Kekule bond-density-wave (BDW) order parameter, which may be favored by the electron-phonon coupling \cite{nomura, mudry}. The internal structure of the vortex in either of the two order parameters in vanishing magnetic field has been a subject of several investigations in the recent past \cite{nomura, mudry, herbut12, hou, bergman}.

    Here I consider the problem of zero-energy states of the two-dimensional Dirac Hamiltonian with a unit vortex in the mass-term, in the presence of either pseudo or true magnetic field. There are two principal results. For the pseudo-magnetic field, a sufficiently extended field's configuration may change the number of zero-energy states to zero or two, depending on the field's direction. In case of a finite pseudo-magnetic flux, on the other hand, the number of zero-energy states remains at unity, in agreement with previous work \cite{pi}.  For true magnetic field, the Dirac equation at zero-energy is shown to be mathematically equivalent to classical Hamilton's equations in a certain time-dependent potential. This analogy is used to show that there is always precisely one zero-energy state irrespectively of the field's profile. This agrees with the solution of the special case of a uniform mass and a field \cite{mudry}. The resulting zero-energy state is naturally time-reversal asymmetric, with the asymmetry growing with the magnetic field. Finally, it is argued that the vortex in a general U(1) order parameter in graphene may carry the electrical charge of zero or one, depending on the type of orders both outside and inside the vortex core. I close with the short discussion of the recent experiment in light of these results.

\section{Equation for zero-energy states}

We are interested in the spectrum of the general two-dimensional Dirac Hamiltonian in presence of the vortex configuration in the two-component mass and the vector potential:
\begin{equation}
\hat{H}= i \gamma_0 \gamma_i (\hat{p}_i - C A_i (\vec{r}) ) - \Delta_1  (\vec{r}) i \gamma_0 \gamma_3 - \Delta_2 (\vec{r})  i \gamma_0 \gamma_5,
\end{equation}
where $i=1,2$, the summation over the repeated indices is assumed, and the matrix $C \in \{I, \gamma_{35}\}$, with $\gamma_{35}=i\gamma_3 \gamma_5$. We chose the  vortex configuration to be rotationally symmetric, so that  $\Delta_1  (\vec{r})= |\Delta(r)| \cos\theta$,  $\Delta_2 (\vec{r})= |\Delta(r)| \sin \theta$, where $\{ r,\theta\}$ are the polar coordinates. The $\gamma$-matrices, as usual, form a representation of the Clifford algebra, and $[\gamma_i, \gamma_j ]_+= 2\delta_{ij}$, for $i=0,1,2,3,5$.

Although the results will be representation independent, to be specific we will use the representation directly relevant to graphene \cite{herbut3}, in which $ i \gamma_0 \gamma_1 =  -\sigma_3 \otimes \sigma_1$,  $ i \gamma_0 \gamma_2 =  -I_2 \otimes \sigma_2$, $ i \gamma_0 \gamma_3 = -\sigma_1 \otimes \sigma_1$, and $ i \gamma_0 \gamma_5 =\sigma_2 \otimes \sigma_1$. $ \{I_2, \vec{\sigma} \}$  is the Pauli basis. The first Pauli matrix acts in the valley, and the second in the sublattice space. In this representation $\gamma_0= I_2\otimes \sigma_3$, and it  anticommutes with the Hamiltonian. In graphene, the Hamiltonian in this representation would act on the Dirac fermion $\Psi = ( u_+, v_+, u_-, v_- )^\top $, with $u_\pm$ ($v_\pm$) representing the wavefunction on sublattice A (B) with Fourier components in the vicinity of the Dirac point $\pm \vec{K}$. When $C=I$ the vector potential represents the true, time-reversal symmetry breaking magnetic field, and when $C=\gamma_{35}$, it stands for the time-reversal preserving pseudo-magnetic field, which would represent the low-energy effect of ripples of the graphene sheet, for example \cite{vafek}. Note that in the above representation $\gamma_{35}= \sigma_3 \otimes I_2$, so that in the latter case the sign of the coupling of the magnetic field to Dirac fermion is different for the two Dirac points.

  The specific mass-term in the graphene representation presented above would describe the vortex configuration in the BDW. The spectrum is nevertheless completely general, as any Hamiltonian with the form of Eq. (1) can be transformed into the graphene representation by a unitary transformation \cite{herbut1}. This follows from the fact that the four anticommuting matrices appearing in the Hamiltonian form a representation of the Clifford algebra, and all such representations are equivalent \cite{schweber}.

  Since the zero-energy states are the eigenstates of $\gamma_0$, it is useful to transform  to the representation in which $\gamma_0$ would become the block-diagonal matrix $\sigma_3 \otimes I_2$. In the graphene representation this is easily accomplished by exchanging the two sets of Pauli matrices.
 The Hamiltonian then becomes purely off-diagonal: $\hat{H}\rightarrow (\sigma_+ \otimes \hat{H}_v + \sigma_- \otimes \hat{H}_u)/2 $, with $\sigma_\pm = \sigma_1\pm i \sigma_2$,  and the equations for the zero-energy states are:
  \begin{equation}
  i (2 \partial_{\bar{z}} + \frac{\chi(r)}{r} z) u_+ (\vec{r}) + \Delta(\vec{r}) u_- (\vec{r}) =0,
  \end{equation}
  \begin{equation}
  - i (2 \partial_{z } - s  \frac{\chi(r)}{r} \bar{z} ) u_- (\vec{r}) + \bar{\Delta}(\vec{r}) u_+ (\vec{r}) =0,
  \end{equation}
  with $v_+=v_- =0$. Here, $z=x+iy$,  $\partial_x +i \partial _y = 2\partial_{\bar{z}}$, and $\Delta= \Delta_1 + i \Delta_2$. We are using the symmetric gauge in which $\vec{A}(\vec{r})  = \chi(r) (-\hat{y},\hat{x})$, so that the field strength is rotationally symmetric and determined by $B(r) = (\partial_r + (1/r))\chi(r)$. The sign $s=1$ refers to the real, and $s=-1$ to the pseudo-magnetic field.

    Note that the other possibility, namely $u_\pm =0$, $\hat{H}_v v=0$, does not yield a normalizable zero-energy state. This is because assuming a finite, real or pseudo, magnetic field at the origin, $\chi(r) \propto r$, and it can be neglected. This implies that $v(r)\propto 1/r$ near the origin, just as without the magnetic field, and one must choose $v_\pm (r)\equiv 0$. For the antivortex the role of $u$ and $v$ are reversed.

 Rotating the variables and assuming rotational symmetry of the solution, as $u_+(\vec{r}) = \sqrt{-i} q(r)$ and $u_- (\vec{r}) = \sqrt{i} p(r)$,  we may finally write the equations as
 \begin{equation}
 \partial_r q(r) = -|\Delta(r)| p (r)-  \chi (r) q (r),
 \end{equation}
  \begin{equation}
 \partial_r p (r)= -|\Delta(r)| q (r) + s \chi (r) p (r),
 \end{equation}
 which are solved in the remainder of the paper.

 \section{Pseudo-magnetic field}

 Consider the simpler case of the pseudo-magnetic field with $s=-1$ first. By rotating in the $(q,p)$-plane by $\pi/4$ it immediately follows that the general solution is
 \begin{equation}
 p (r) \pm  q  (r) =  A_\pm  e^{-\int_0 ^r (\pm |\Delta(r')| + \chi(r')) dr'}.
 \end{equation}
 Assuming that the amplitude $|\Delta(r\rightarrow \infty )| \rightarrow \Delta_0$, when $\chi(r)=0$  normalizability of the state dictates that the constant $A_-=0$, and therefore $p(r)=q(r)$. In this case there is a single zero-energy state \cite{jackiw}. If $\chi(r)\neq 0$, but $|\chi(r\rightarrow\infty)| < \Delta_0$, it is still $A_-=0$, and there is still a single zero-energy state. If $\chi(r)<0$, but $|\chi(r\rightarrow\infty)| > \Delta_0$, there are no normalizable zero-energy states at all, as both constants $A_\pm=0$.

 When $\chi(r)>0$, and $\chi(r\rightarrow \infty )\sim r^\alpha$ with $\alpha > 0$, on the other hand, there are evidently {\it two} orthogonal normalizable zero-energy states. We may write them as
 \begin{equation}
 u^\dagger _\pm = {\cal N}_\pm  e^{-\int_0 ^r ( \pm |\Delta(r')| + \chi(r'))dr' } (\pm 1, i),
 \end{equation}
 where ${\cal N}_\pm$ are the  normalization constants. The norm of both states will be finite if the strength of the field $B(r)$ decays slower with radius than $1/r$. In particular, if the pseudo-magnetic field and the amplitude of the mass are uniform, whereas the first state is centered at the origin the second state is sharply peaked at $r_{max} = 2 \Delta_0/B$, which diverges as the field approaches zero.

   In sum, for a sufficiently localized pseudo-magnetic field of either sign, there is precisely  one zero-energy state localized at the origin \cite{pi}. If the field decreases slower than the inverse radius, on the other hand, the number of zero-energy states is zero or two, depending on the field's direction.

\section{Magnetic field: analogy with mechanics}

   For the true magnetic field ($s=1$), assuming  $|\Delta(r)|$ to be monotonically increasing, we may  introduce a time-like dimensionless variable $t$ as $dt= |\Delta(r)| dr$, and rewrite Eqs. (4)-(5) in a more suggestive form as
   \begin{equation}
   \dot{q} (t) = -f(t) q(t) -p (t),
   \end{equation}
   \begin{equation}
   \dot{p} (t) = -q(t) + f(t) p (t),
   \end{equation}
   where $\dot{x}= dx/dt$, and $f(t) = \chi(t)/|\Delta(t)|$. In this form the Dirac equation for the zero-energy state may be recognized as Hamilton's equations of classical mechanics for the "coordinate" $q$ and the "canonical momentum" $p$, with the time-dependent classical Hamilton's function
   \begin{equation}
   H(q,p,t) = -f(t) p q +\frac{1}{2}(q^2 - p^2),
   \end{equation}
    and with the corresponding Lagrangian
    \begin{equation}
    -L (q,\dot{q},t) = \frac{1}{2} \dot{q}^2 + \frac{\kappa(t) }{2} q^2 ,
    \end{equation}
    with $\kappa(t)= 1+ f(t)^2 - \dot{f} (t)$, and with a total time derivative omitted. Apart from the irrelevant minus sign, this Lagrangian describes the motion of a classical particle in the inverted harmonic potential
    \begin{equation}
    V(q,t) = - \frac{ \kappa(t)}{2}  q^2.
    \end{equation}

 In absence of the magnetic field $\kappa(t)\equiv 1$,  and the potential $V(q,t)$ is  static; for any initial $q(0)$ providing the right amount of initial kinetic energy will get the particle to the origin in infinite time. Since the initial energy is simply $E (t=0) = (p^2 (0) - q^2(0))/2$, this is ensured by the initial condition $p(0)=q(0)$. The conservation of energy implies then that $p(t)=q(t)$ at all times. This mechanical interpretation of the familiar solution becomes useful in understanding the qualitative effect of the magnetic field. Although for a finite magnetic field the potential $V(q,t)$ typically becomes steeper with time, it is clear that there are two independent solutions of the equation of motion: first when the initial kinetic energy is too small so that particle starting from some $q(0) >0$ returns and ultimately runs off to positive infinity, and the second, when for a too large initial kinetic energy the particle goes over the top, and runs off to the negative infinity. Continuity guarantees then the existence of the initial condition in between these two extremes for which the particle will reach the top with precisely zero velocity, i. e. in infinite time. This trajectory corresponds to the normalizable zero-energy state of the Dirac Hamiltonian. Using the equation of motion the rate of the change in the "mechanical energy" of the system is
 found to be
 \begin{equation}
      \dot{E}= -\frac{\dot{\kappa}(t) }{2} q^2 (t).
 \end{equation}
 To reach the top with zero velocity the sum of the initial energy and the work done externally needs to vanish. This may be expressed as a global condition on the solution,
 \begin{equation}
      p^2 (0) + 2 \int_0 ^\infty \kappa (t) q(t) \dot{q} (t) dt=0.
 \end{equation}

\section{Example: uniform magnetic field }

 For illustration, let us consider the case of the uniform magnetic field $\chi(r) = Br/2$, and the vortex with the core of size $R$: $|\Delta(r)| = \Delta_0 r/R$, for $r<R$, and  $|\Delta(r)| = \Delta_0$, for $r\geq R$.  The equation of motion is
 \begin{equation}
         \ddot{q}=(1+ f^2(t) -\dot{f}(t)) q.
         \end{equation}
 For $r<R$, $f(t)= BR / 2\Delta_0$ and constant, and $t= \Delta_0 r^2 /2R$. The general solution is therefore
         \begin{equation}
         q_< (r)= c_1 e^{\frac{r^2}{2L^2}} + c_2 e^{- \frac{r^2}{2L^2}},
         \end{equation}
 where $c_1$ and $c_2$ are constants, and $L$ is the characteristic length, determined by
         \begin{equation}
         L^{-4}= (\frac{B}{2})^2 + (\frac{\Delta_0}{R})^2.
         \end{equation}

    When $r>R$, on the other hand, $f= \delta t$, where $\delta= B/(2(\Delta_0)^2 ) $ is a dimensionless parameter. It would then seem that one needs to distinguish two cases:

\noindent
    a) when  $\delta <1$,
    \begin{equation}
    q_> (r) = c_3 ( \Delta_0 r ) e^{ -\frac{|B| r^2}{2}} U[ \frac{3|\delta| + 1- \delta}{4|\delta|}, \frac{3}{2}, \frac{|B|r^2}{2}],
    \end{equation}
    where U stands for the hypergeometric function. In this case the potential $V(q,t)$ is always repulsive and consequently the solution is monotonically decreasing.

\noindent
    b) when $\delta > 1$, the solution is
    \begin{equation}
    q_> (r) = c_3 ( \Delta_0 r ) e^{ -\frac{|B| r^2}{2}} U[ \frac{1+2\delta}{4\delta}, \frac{3}{2}, \frac{|B|r^2}{2}].
    \end{equation}
    Although the potential $V(q,t)$ now starts as attractive, the solution is nevertheless still always monotonically decreasing, and without oscillations. It is in fact qualitatively the same as in the previous case.

    The continuity implies that $q_< (R)= q_> (R)$ and $dq_< (R)/dr = dq_>(R)/dr$, which yields two linear equations on the three constants $c_i$, $i=1,2,3$. The normalization then provides the third condition that completes the solution, as usual. It is also easy to see that the solution is always with $c_2 > c_1 $ and therefore monotonically decreasing, as one indeed would expect from the mechanical analogy.

\section{Remarks on the solution}

      The reversal of the sign of the true magnetic field simply exchanges the solutions for $p$ and $q$, as evident from Eqs. (8)-(9). This is of course equivalent to the time reversal, consisting of the exchange of the two Dirac points followed by the complex conjugation \cite{herbut3}.

 The solution for the antivortex may be obtained most simply by multiplying the zero-energy state by the matrix $\gamma_5$, for example. Since $\gamma_5$ anticommutes with the last term in Eq. (1) while commuting with the rest, the Hamiltonian $\gamma_5 \hat{H} \gamma_5$ has only the sign of $\Delta_2 (\vec{r})$ flipped, and thus represents an antivortex. In the graphene representation $\gamma_5= \sigma_2 \otimes\sigma_2$, and the result of the multiplication correctly reproduces the time reversal  followed by the exchange of the sublattices.

   It is interesting to consider the  two limits of the above zero-energy solution, for weak and strong magnetic field. In the former case the zero-energy state approaches the standard zero-field solution of Jackiw and Rossi which respects the time-reversal symmetry.  In the latter case, when $B\gg \Delta_0/R$, the solution for $q(r)$ is localized very close to the center of the core, and thus it has the form of the second term in Eq. (16). Eq. (4) then fixes the second component to be $p(r) \sim ( BR/\Delta_0 ) q(r) \gg q(r)$. In the limit of strong magnetic field the zero-energy solution is, in the graphene representation,
   \begin{equation}
   \Psi_0 ^\dagger = {\cal N} (0, 0, e^{-r^2/4l^2},0)+ O(\Delta_0/BR),
   \end{equation}
   and has a single large component. Note how the result does not approach continuously the solution at $\Delta=0$, since even for an infinitesimal $\Delta$ vorticity eliminates the $v$-components of the state completely. When the  magnetic field is much larger than the mass the zero-energy state is strongly localized, finite only over one sublattice, and has Fourier components essentially only near one of the Dirac points.

   Since the Hamiltonian in Eq. (1) always anticommutes with $\gamma_0$, even a rotationally asymmetric mass-vortex and/or the magnetic field still produce a spectrum symmetric around zero. This means that continuous deformations of the rotationally invariant problem considered here would still yield exactly one zero-energy state, because moving that single state to a finite energy would obviously violate the spectrum's symmetry.

\section{Electrical charge of the vortex}

   The existence of the zero-energy solutions in the magnetic field being established, let us turn to some of their possible manifestations. Let us assume first that the ordered state of the electrons in the magnetic field at half filling, i. e. at the filling factor zero, preferred by the electron-electron interactions is indeed the BDW \cite{nomura, mudry}, identical for both projections of the spin, for example. This being an order parameter with an (approximate) U(1) symmetry, a vortex will be topologically distinct and thus stable. The mean-field Hamiltonian in presence of such a vortex will then consist of two (spin) copies of the Eq. (1) in the true magnetic field, which will have two zero-energy states. The zero-energy Hilbert space then provides a representation of the algebra of the possible order parameters in the core \cite{herbut1}. In the  case at hand this algebra consists of the arbitrary oriented N\' eel and the charge-density-wave (CDW) order parameters, given by $\{ \vec{\sigma}\otimes \gamma_0, I_2 \otimes \gamma_0 \}$, respectively, since these four matrices exhaust the set of operators that anticommute with the Hamiltonian $I_2 \otimes \hat{H}$. The electrical charge of the vortex vanishes if the N\' eel order parameter is preferred locally, since that implies that precisely one state from the zero-energy space is occupied and the other orthogonal state is empty.  If, in contrast, the CDW state is preferred energetically, both states will be occupied (for vortex) or empty (for antivortex) and the vortex will bind a unit charge \cite{hou}. Obviously there is a cost in the  electrostatic energy in the latter case, but one can conceive of, possibly unrealistic, Hamiltonians, with the nearest-neighbor sufficiently strong, for example \cite{herbut3}, where such a state could be energetically favorable. At finite temperatures, fluctuating vortices and antivortices  would in that case indeed behave like a plasma of positive and negative charges, with the metallic conductivity proportional to the vortex density. Assuming that density vanishing at the point of metal-insulator transition in the standard Kosterlitz-Thouless fashion would then rationalize the experimental observation  \cite{ong}. This scenario, however, obviously depends on the favorable ordering of the interaction energy scales in the many-electron Hamiltonian, so that electron-phonon coupling is larger than nearest-neighbor repulsion, which in turn is larger than the on-site repulsion, for example. Recent work \cite{weeks} which finds the stable BDW  over a large portion of the zero-field phase diagram, however, could make this mechanism more plausible.

 The second option mentioned in the introduction for the order parameter with the (exact) U(1) symmetry is the familiar N\' eel order, which for small enough Zeeman energy develops an easy-plane \cite{herbut2}.  Vortex in this order would be described by the mass-term in the Eq. (1) replaced by $(\Delta_1 \sigma_1 + \Delta_2 \sigma_2)  \otimes \gamma_0$. This being equivalent to Eq. (1), the Hamiltonian for the N\' eel vortex also contains two zero-energy states in its spectrum \cite{herbut12}. This time however, the algebra of competing orders in the vortex core is different, and is readily seen to consist of the third component of the N\' eel, two BDWs, and the third component of the Haldane-Kane-Melle (HKM) order parameter
 \cite{haldane}, $\{ \sigma_3 \otimes \gamma_0, I_2 \otimes i\gamma_0 \gamma_3, I_2 \otimes  i\gamma_0 \gamma_5, \sigma_3 \otimes i\gamma_1 \gamma_2 \}$, respectively. In this case the vortex will be charged {\it only} if it is the HKM ordering that is energetically preferred in the core, and otherwise not. In particular, one would assume that the order in the core, if possible, should be of the same type as the order supporting the vortex, which was preferred energetically in the first place. On this basis it is the third component of the N\' eel order that may be expected to develop in the core \cite{herbut12}, with the consequence of rendering the vortex electrically neutral.

 \section{Conclusion}

 In this paper it has been established that the two-dimensional Dirac equation with a vortex in the two-component mass-term has a single zero-energy solution even in presence of an arbitrary, including infinite, flux of the (true)  magnetic field. When the particle in question has spin-1/2, the number of zero-modes is doubled. In the context of graphene this should imply that, depending on the residual terms in the Hamiltonian such as electron-electron interactions, non-linearities of the electron dispersion, edges, etc., the electrical charge of the  vortex excitation in some $U(1)$ order parameter in magnetic field may be zero or one, corresponding to the occupancy of both, or just one of the two zero-energy modes. Two candidate order parameters and possible connections to the experiment on metal-insulator transition in graphene were discussed.

 \section{Acknowledgement}

 This work is supported by the NSERC of Canada. I am grateful to R. Jackiw, V. Juri\v ci\'c, B. Roy, G. Semenoff, and B. Seradjeh for useful discussions,  and the Aspen Center for Physics for hospitality.

\end{document}